\documentclass[aps,10pt,twocolumn,tightenlines,nofootinbib,floatfix]{revtex4-1}
\usepackage[utf8]{inputenc}
\usepackage{amsmath}
\usepackage{appendix}
\usepackage{natbib}
\usepackage{graphicx}
\usepackage{geometry}
\usepackage{ragged2e}
\usepackage{amsfonts,amssymb}
\usepackage[table]{xcolor}
\usepackage{array}
\usepackage[normalem]{ulem}
\newcolumntype{C}[1]{>{\centering\let\newline\\\arraybackslash\hspace{0pt}}m{#1}}

\newcommand{\lsim}{\mathrel{\hbox{\rlap{\lower.75ex \hbox{$\sim$}} \kern-.3em \raise.4ex \hbox{$<$}}}}
\newcommand{\gsim}{\mathrel{\hbox{\rlap{\lower.75ex \hbox{$\sim$}} \kern-.3em \raise.4ex \hbox{$>$}}}}

\begin{document}

\title{Anomalous ANITA air shower events and tau decays}

\author{Shoshana Chipman}
\affiliation{Department of Astronomy and Astrophysics, University of Chicago, Chicago, IL 60637, USA}
\author{Rebecca Diesing}
\affiliation{Department of Astronomy and Astrophysics, University of Chicago, Chicago, IL 60637, USA}
\author{Mary Hall Reno}
\affiliation{Department of Physics and Astronomy, University of Iowa, Iowa City, IA 52242, USA}
\author{Ina Sarcevic}
\affiliation{Department of Physics, University of Arizona, 1118 E. 4th St. Tucson, AZ 85704, USA}

\date{June 27, 2019}

\begin{abstract}
Two unusual neutrino events in the Antarctic Impulse Transient Antenna (ANITA) appear to have been generated by air showers from a particle emerging from the Earth at angle $\sim 25^\circ-35^\circ$ above the horizon. We evaluate the effective aperture for ANITA with a simplified detection model to illustrate the features of the angular dependence of expected events for incident standard model tau neutrinos and for sterile neutrinos that mix with tau neutrinos.
We apply our sterile neutrino aperture results to a dark matter scenario with long-lived supermassive dark matter that decay to sterile neutrino-like particles.
We find that for up-going air showers from tau decays, from isotropic fluxes of standard model, sterile neutrinos or other particles that couple to the tau through suppressed weak interaction cross sections cannot be responsible for the unusual events. 
\end{abstract}

\maketitle

\section{Introduction}

Neutrinos of astrophysical origin present the opportunity to explore and understand the conditions of cosmic ray acceleration and the surrounding astrophysical environment \cite{Kotera:2011cp,Anchordoqui:2018qom}. 
A number of detectors, current and proposed, rely on
neutrino interactions in water or ice. Detectors include IceCube \cite{Aartsen:2018vtx}, ANTARES
\cite{Albert:2017nsd}, KM3net \cite{Sanguineti:2019pkv}, ARA \cite{Allison:2015eky}  and ARIANNA \cite{Anker:2019mnx}). Air shower  signals via particles, fluorescence, radio and optical Cherenkov are the target of surface instruments such as Auger \cite{Aab:2015kma,Aab:2019gra,Zas:2017xdj}, the
Telescope Array \cite{Abbasi:2019fmh},  MAGIC \cite{Aartsen:2018vtx}, GRAND \cite{Alvarez-Muniz:2018bhp}, and Trinity \cite{Otte:2018uxj}. 
Above the Earth, proposed satellite-based instruments
sensitive to upward-going air showers include CHANT \cite{Neronov:2016zou} and POEMMA \cite{Olinto:2017xbi}. The balloon-borne ANITA detector
\cite{Allison:2018cxu,Gorham:2019guw,Gorham:2016zah,Gorham2018} is sensitive to neutrino interactions 
in the Antarctic ice where the Askaryan effect is important. ANITA is also sensitive to tau neutrino charged current interactions in the Earth that produce taus that decay in the atmosphere.  These upward-going tau shower signals 
at ANITA are the focus of this paper.

For neutrino telescopes, the standard model source of taus is tau neutrino charged-current interactions in the Earth. 
With neutrinos coming from charged pion decays and nearly bi-maximal $\nu_\mu-\nu_\tau$ mixing, over astronomical distances approximately equal fluxes of electron neutrinos, muon neutrinos and tau neutrinos
arrive at the Earth \cite{Learned:1994wg}. Lepton flavor universality has the three standard model neutrinos with equal interaction cross sections on nucleon targets and, at high energies, the neutrino and antineutrino cross sections are equal \cite{Gandhi:1995tf,Gandhi:1998ri,Jeong:2010za,Connolly:2011vc,CooperSarkar:2011pa}. Below, ``neutrino'' refers to both particle and antiparticle.

The ANITA collaboration has reported observations of two unusual events are consistent with shower characteristics of upward-going taus that decay in the atmosphere \cite{Gorham:2016zah,Gorham2018}. In the ANTIA-III run, event 15717147 had an estimated shower energy of $0.56^{0.3}_{-0.2} \times 10^{9}$ GeV, and emerged with an azimuthal elevation angle of $-35.0^\circ \pm 0.3^\circ$ -- which is to say that the event emerged at around $35^\circ$ above the horizon, or around $55^\circ$ from the vertical \cite{Gorham2018}. Additionally, the first run of ANITA in 2016 run produced event 3985267, of shower energy of  $(0.6 \pm 0.4) \times 10^9$ GeV, which emerged at an azimuthal elevation angle of $-27.4^\circ \pm 0.3^\circ$, roughly $63^\circ$ from the zenith \cite{Gorham:2016zah}.
The interpretation of these unusual events as coming from tau neutrinos is problematic \cite{Gorham:2016zah,Gorham2018,Romero-Wolf:2018zxt} because of the energies of the showers and the apparent angles of the tau neutrinos that induced them.

One challenge to the tau neutrino interpretation is
neutrino flux attenuation.
While governed by weak interactions, 
the neutrino interaction length (in units of column
depth)
$\lambda_\nu=(N_A \sigma _{\nu N})^{-1}$
\cite{Gandhi:1995tf,Gandhi:1998ri,Jeong:2010za,Connolly:2011vc,CooperSarkar:2011pa} is large compared 
to the column depths traversed by the neutrino trajectories of the unusual events. As an indication
of the scales involved, for example,
for a neutrino incident at nadir angle $0^\circ$, the column depth is $\sim 1.1\times 10^{10}$ g/cm$^2$, equal to the neutrino interaction length for $E_\nu\sim 40$ TeV. 
For a nadir angle of $60^\circ$ (elevation angle $40^\circ$), the neutrino interaction length equals the column depth in Earth when $E_\nu\sim 250$ TeV. 

As the neutrino energy increases, the effective solid angle that can be detected decreases. The neutrino fluxes incident at small nadir angles, or alternatively, emerging at large elevation angles, can be
significantly attenuated. At the elevation angles of $25-35^\circ$ of the high energy ANITA events, the tau exit probability is small in the standard model.  Additionally, it is difficult to explain why events are detected at large elevation angles but not small
elevation angles where the exit probabilities are larger \cite{Romero-Wolf:2018zxt,Fox:2018syq}.

Physical effects related to the ice/air boundary for downward-going cosmic ray air showers are under discussion as possible explanations of ANITA's unusual events \cite{deVries:2019gzs,Shoemaker:2019xlt}. 
For example, explanations point towards the Antarctic subsurfaces and 
firn density inversions as well as the ice structure as a possible 
explanation \cite{Shoemaker:2019xlt}. A beyond the standard model (BSM) explanation proposed for downward-going air showers is axion-photon conversion
\cite{Esteban:2019hcm}.

There are also a number of BSM physics explanations
for upward-going air showers that come from tau decays or other particle decays in the atmosphere.
Neutrino production of heavy BSM particles that decay directly to taus 
or to other BSM particles that couple to taus have been introduced modify the standard model large elevation angle suppression \cite{Connolly:2018ewv,Fox:2018syq,Collins:2018jpg,Chauhan:2018lnq,Anchordoqui:2018ssd}.
Several scenarios with decaying heavy dark matter have been
proposed, including the one in which decaying dark matter is
trapped in the Earth \cite{Anchordoqui:2018ucj}, and
others
in which the dark matter decays in the galactic halo
that ultimately produce shower \cite{Heurtier:2019git,Heurtier:2019rkz,Cline:2019snp} or Askaryan events \cite{Hooper:2019ytr} in ANITA. 
Sterile neutrinos that interact to produce taus have been proposed to avoid the neutrino flux attenuation
at large elevation angles \cite{Huang:2018als,Cherry:2018rxj}.

In general, the ANITA events are in 
tension with other constraints, for example, as
discussed in Ref. \cite{Cline:2019snp}.
While there are scenarios that may be acceptable, e.g.,
boosted 
dark matter decays into lighter dark matter which decays into hadrons for specific model parameters \cite{Heurtier:2019rkz}, it is a challenge to
describe the ANITA unusual events including the
emergence angles and not over predict
IceCube and Auger event rates. 
 
Most of the BSM analyses use approximate analytic results and a narrow energy range associated with ANITA's unusual events. In this paper, we
consider a range of energies and angles using 
Monte Carlo simulations of neutral particle interactions
that couple to taus, and a stochastic evaluation of tau energy loss in the Earth \cite{Reno:2019jtr}. Using a simplified model of the ANITA detection probability, we find the angular dependence of the effective aperture.
Our analysis allows us to separate the particle physics effects (both standard model and a sterile neutrino example of BSM physics) from the tau air shower, detection and surface geometry effects in the evaluation of the effective aperture.

We start with the standard model evaluation of the tau exit probabilities. We also consider a modification of the standard model tau neutrino cross section with a suppression associated with a color glass condensate treatment of the high energy extrapolation of the neutrino-nucleon cross section \cite{McLerran:1993ni,JalilianMarian:1996xn,McLerran:1998nk,Kovchegov:1996ty,Kovchegov:1997pc,Henley:2005ms}.
The ``sterile neutrino'' consider here is a generic neutral BSM particle with suppressed cross sections, assumed to couple to taus by charged current interactions with a cross section
$\sigma_{\nu_s N}=\epsilon_\nu\sigma_{\nu N}$.

We conclude that standard model $\nu_\tau$'s from an isotropic flux cannot account for the unusual events. Our results are consistent with those of Ref. \cite{Romero-Wolf:2018zxt} and others \cite{Shoemaker:2019xlt,Cline:2019snp}.
Our quantitative evaluation of the exit probabilities for $\tau$'s from sterile neutrino interactions in the Earth demonstrates that even with no flux attenuation in the Earth, the lack of events at lower elevation angles makes even a large isotropic flux of sterile neutrinos a poor candidate source of the ANITA unusual events. 
We find that in principle an energy threshold effect can enhance
large elevation angle events relative to small angles. 
A mono-energetic source in the energy threshold region that may produce upward-going air showers, the feebly interacting $\chi$ from supermassive dark matter in the model of Hooper et al. in Ref. \cite{Hooper:2019ytr},
is used as an example to demonstrate this effect.

In the next section, we outline our approximate evaluation of the effective aperture for standard model neutrinos and for sterile neutrinos with suppressed cross sections.  We discuss the geometric and neutrino interaction origins of the angular distribution of the effective aperture. In Sec. III, we discuss how signals at large elevation angles may be enhanced by showers with energies near the ANITA energy threshold. We demonstrate the effect with a sterile neutrino example and $\sim 400$ PeV supermassive dark matter decays in the galactic halo \cite{Hooper:2019ytr}. 
A related IceCube signal is a constraining feature. Finally, we summarize our results in Sec. IV.

\section{Effective Aperture}

\begin{figure}
\includegraphics[width=0.75\columnwidth, trim=6cm 8cm 5cm 7cm, clip]{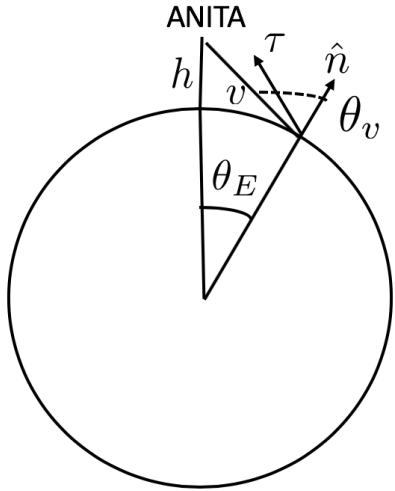}
\caption{Geometry of ANITA an altitude $h$ above the surface of the Earth. The line of sight from the tau exit point at a co-latitude $\theta_E$ has length $v$ and makes an angle $\theta_v$ relative to the local normal $\hat{n}$.}
\label{fig:geometry-overview}
\end{figure}
\begin{figure}
\includegraphics[width=0.8\columnwidth, trim=6cm 11cm 5cm 11cm, clip]{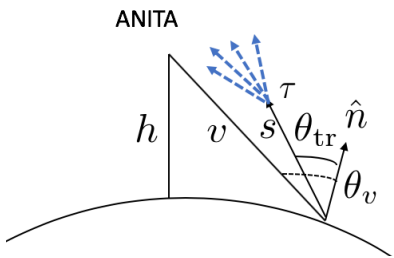}
\caption{The tau trajectory make an angle $\theta_{\rm tr}$ relative to the local normal $\hat{n}$. The tau decays a distance $s$ along its trajectory. The figure is exaggerated to distinguish  $\theta_v$ and $\theta_{\rm tr}$, however the effective Cherenkov angle of the signal from the tau decay shower is such that $\theta_v\simeq\theta_{\rm tr}$.}
\label{fig:geometry}
\end{figure} 

\subsection{Overview}

The first step in determining an event rate is finding the effective aperture for ANITA.
The ANITA effective
aperture $\langle A\Omega\rangle$ 
depends on the viewable area of the Earth below the detector and on the probability to observe a tau decay induced air shower at a given angle and altitude. The geometry is illustrated
in Figs. \ref{fig:geometry-overview} and
\ref{fig:geometry}. The ANITA detector is
at altitude $h$, taken to be $h=35$ km
\cite{Romero-Wolf:2018zxt}.
The signals considered here come from air showers along the
trajectory of an emerging tau that makes an angle
$\theta_{\rm tr}$ with respect to the local 
normal $\hat{n}$ at its point of emergence from the Earth, which is at a co-latitude $\theta_E$ relative to the line from the center of the Earth to ANITA.
The viewing angle $\theta_v$ is the angle from ANITA to the point at which the tau emerges. 
It is convenient to describe elements of the observation probability in terms of the tau elevation
angle $\beta_{\rm tr}$, related to the trajectory's angle to $\hat{n}$ by
\begin{equation}
    \beta_{\rm tr}+\theta_{\rm tr} = \pi/2\ .
\end{equation}

The effective aperture for incident tau neutrinos of energy $E_{\nu_\tau}$ can be written as
\cite{Motloch:2013kva}, 
\begin{equation}\label{eq:apereqn}
\left<A\Omega\left(E_{\nu_{\tau}}\right)\right> = \int_{S}\int_{\Delta\Omega_{\rm tr}} P_{\rm obs}\ \hat{r}\cdot\hat{n}\, dS \, d\Omega_{\rm tr}\ ,
\end{equation}
where $\hat{r}\cdot\hat{n} = \cos\theta_{\rm tr }$, $P_{\rm obs}$ is the detection probability, and  
$dS$ is the area element on the surface of the Earth. The integral $d\Omega_{\rm tr}$ accounts for the trajectories of the tau for which the air shower is detected.

The
effective Cherenkov angle is $\theta^{\rm eff}_{\rm Ch}\sim 1^\circ$. 
In all that follows, we approximate $\theta_{\rm tr}\simeq\theta_v$ since Cherenkov angle is small. This simplifies the evaluation of the effective aperture. The effective aperture is then approximately \cite{Motloch:2013kva} (see also, Ref. \cite{Reno:2019jtr}, Appendix A),
\begin{equation}
 \left<A\Omega\left(E_{\nu_{\tau}}\right)\right> \simeq 2\pi^2 R_E^2 \sin^2\theta_{\rm Ch}^{\rm eff}
 \int P_{\rm obs}\, \cos\theta_v\sin\theta_E
 d\theta_E\ .
 \label{eq:Aomega-approx}
\end{equation}
The radius of the Earth is $R_E=6371$ km. 

The probability $P_{\rm obs}$ that a tau neutrino with energy $E_{\nu_{\tau}}$ produces a shower that is detectable is \cite{Romero-Wolf:2017ope,Romero-Wolf:2018zxt}
\begin{align}\label{eqnpobs}
P_{\rm obs} &= \int p_{\rm exit}\left(E_{\tau}|E_{\nu_{\tau}},\theta_{\rm tr}\right) \nonumber \\
& \times \left[\int ds\, {p}_{\rm decay}(s) P_{\rm det}\left(E_{\tau},\theta_{v},\theta_{\rm tr},s\right)\right]\, dE_{\tau}\, .
\end{align}
As noted above, we approximate $\theta_{v}\simeq\theta_{\rm tr}$ in the discussion below.
For $p_{\rm decay}$ and $P_{\rm det}$, the distance $s$ is
the length of the tau path length from its exit point on Earth to its point of decay.

In the next section, we discuss the exit probabilities and emerging tau energies as a function of $\beta_{\rm tr}$ for the standard model and variations. The decay and detection probabilities for tau decays are independent of exit probabilities. The decay probability density is
\begin{equation}
    p_{\rm decay} = \frac{\exp\Bigr(-s/(\gamma c\tau)\Bigl)}
    {\gamma c\tau}
\end{equation}
where $\gamma=E_\tau/m_\tau c^2$ is the usual
gamma-factor of time dilation for tau decays. The decay length of the tau is $\gamma c\tau\simeq 5\ {\rm km}\times (E_\tau/10^8\ {\rm GeV})$.

For the detection probability, we use a simplified model of the ANITA-III detector. ANITA detects the electric field generated by the shower.  We approximately follow Ref. \cite{Connolly:2018ewv}. We take the probability of detection as the product of theta functions times the hadronic branching fraction of the tau $B_{\rm had}=0.648$:
\begin{eqnarray}
\nonumber
    P_{\rm det}(E_\tau,\beta_{\rm tr})
    &=& B_{\rm had}\\ 
    \nonumber
    &\times&\theta\Biggr(\frac{E_{\rm shr}}{10^8\ {\rm GeV}}\frac{74\ {\rm km}}{r_0(E_\tau,s_d,\beta_{\rm tr})}-1\Biggr)\\
    &\times& \theta(6\ {\rm km}-a)\ .
    \label{eq:pdet}
\end{eqnarray}
The first theta function enforces a minimum electric field requirement which to first approximation follows from the fact that 
higher the shower energis have larger the electric fields. Up to a point, the shorter the distance from the start of the shower to ANITA,
$r_0\simeq v - s$, the larger the electric field at the antennas.
The second theta function cuts off the integration over altitude at 6 km.
As the altitude at which a tau shower begins is increased, first the electric field gets larger, but after $\sim 6$ km, the angle between the shower
axis and line of sight ($\theta_{\rm view}$ in
ref. \cite{Romero-Wolf:2018zxt}) decreases \cite{Romero-Wolf:2018zxt}. 

Eq. (\ref{eq:pdet}) is a rough approximation to a more detailed model of the electric field from tau showers \cite{Romero-Wolf:2018zxt}. We show below that using eq. (\ref{eq:pdet}) with $E_{\rm shr}=0.98E_\tau$ as in Ref. \cite{Romero-Wolf:2018zxt} and
\begin{equation}
    \theta_{\rm Ch}^{\rm eff}\simeq 1.0^\circ - 0.02\beta_{\rm tr}\ ,
\end{equation}
for $\beta_{\rm tr}$ in degrees, the mean ANITA-I,III effective aperture is reasonably well
reproduced.

\subsection{Exit probability and effective aperture}

\begin{figure}
\includegraphics[width=\columnwidth]{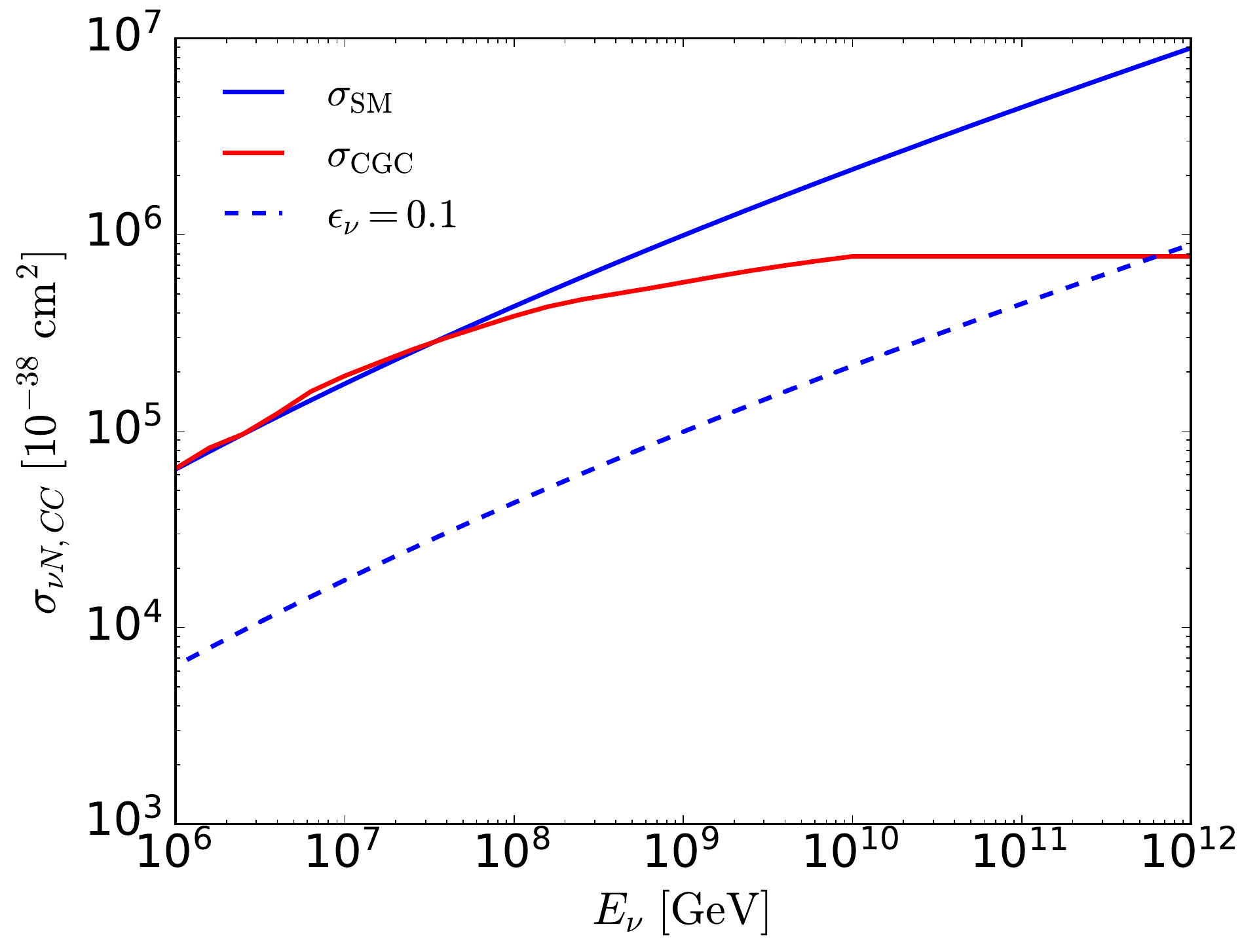}
\caption{The neutrino-nucleon charged current cross section as a function of neutrino energy for an evaluation using CT14 \cite{Dulat:2015mca} parton distribution functions (SM), with color glass condensate suppression at high energies (CGC) \cite{Henley:2005ms} and for $\sigma=\epsilon_\nu\sigma_{\rm SM}$ with $\epsilon_\nu=0.1$.}
\label{fig:sigma}
\end{figure} 

The quantity $p_{\rm exit}\left(E_{\tau}|E_{\nu_{\tau}},\theta_{\rm tr}\right)$ is the 
exit probability density, which depends on the neutrino-nucleon cross section.
The exit probability is
\begin{equation}
    P_{\rm exit }(E_{\nu_\tau}\theta_{\rm tr})= \int dE_\tau \left(E_{\tau}|E_{\nu_{\tau}},\theta_{\rm tr}\right)\ .
\end{equation}
Our standard model cross section for neutrino-isoscalar nucleon scattering is calculated with the CT14 parton distribution functions \cite{Dulat:2015mca}. The cross section is shown with
the solid blue line in Fig. \ref{fig:sigma}.
We use cumulative distribution functions to sample the energy distribution of the taus that exit
for a given incident tau neutrino energy $E_{\nu_\tau}$ and angle $\beta_{\rm tr}$,
as described in detail in Ref. \cite{Reno:2019jtr}.

The exit probability also depends on the tau electromagnetic energy loss.
In charged current scattering, a tau is produced. The tau loses energy primarily through electron-positron pair production and photonuclear interactions, which we implement with a Monte Carlo simulation \cite{Dutta:2000hh} 
that also includes tau neutrino regeneration. We use the Abramowicz et al. (ALLM) parameterization of the electromagnetic structure function $F_2$ \cite{Abramowicz:1991xz,Abramowicz:1997ms} in our evaluation of the photonuclear contribution. At high energies, extrapolations of $F_2$ beyond the measured regime introduce uncertainties. 

A feature unique to tau neutrinos is the significance of tau neutrino regeneration with neutrino interactions in the Earth \cite{Halzen:1998be,Iyer:1999wu,Dutta:2000jv,Becattini:2000fj,Beacom:2001xn,Alvarez-Muniz:2017mpk,Reno:2019jtr}. Tau neutrino regeneration comes from tau neutrino charged-current production of taus which subsequently decay back to $\nu_\tau$.  Through a series of neutrino interaction and decay, high energy tau neutrinos can produce taus that emerge from the Earth to produce up-going air showers  \cite{Domokos:1997ve,Domokos2,Fargion:2000iz,Fargion:2003kn,Fargion:2003ms,Bertou:2001vm,Feng:2001ue,Lachaud:2002sx,Bottai:2002nn,Hou:2002bh,Tseng:2003pn,Aramo:2004pr,Dutta:2005yt,Asaoka:2012em,PalomaresRuiz:2005xw}. More details on the evaluation of the tau exit probabilities appear in Ref.
\cite{Reno:2019jtr}. 

\begin{figure}
\includegraphics[width=\columnwidth]{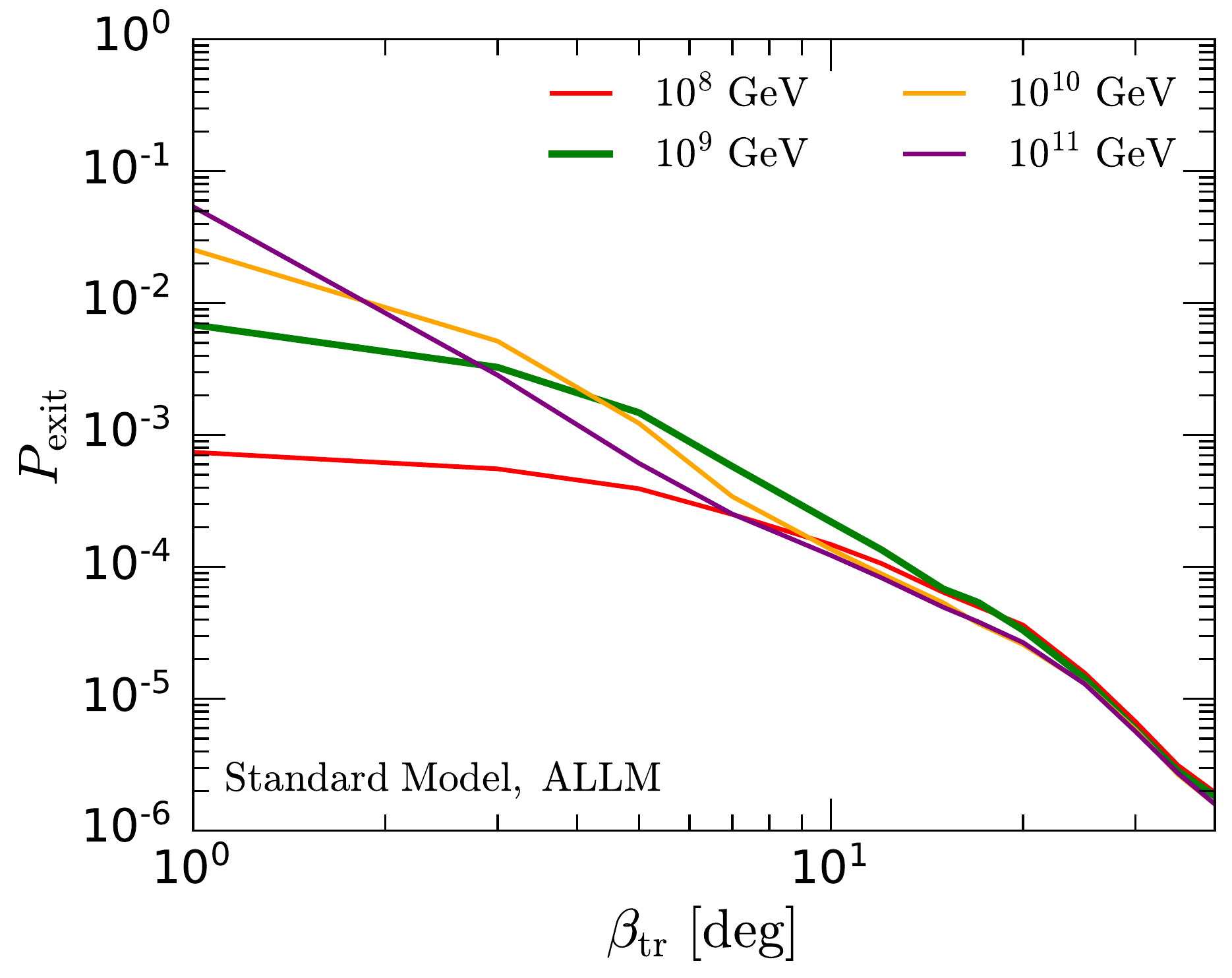}
\includegraphics[width=\columnwidth]{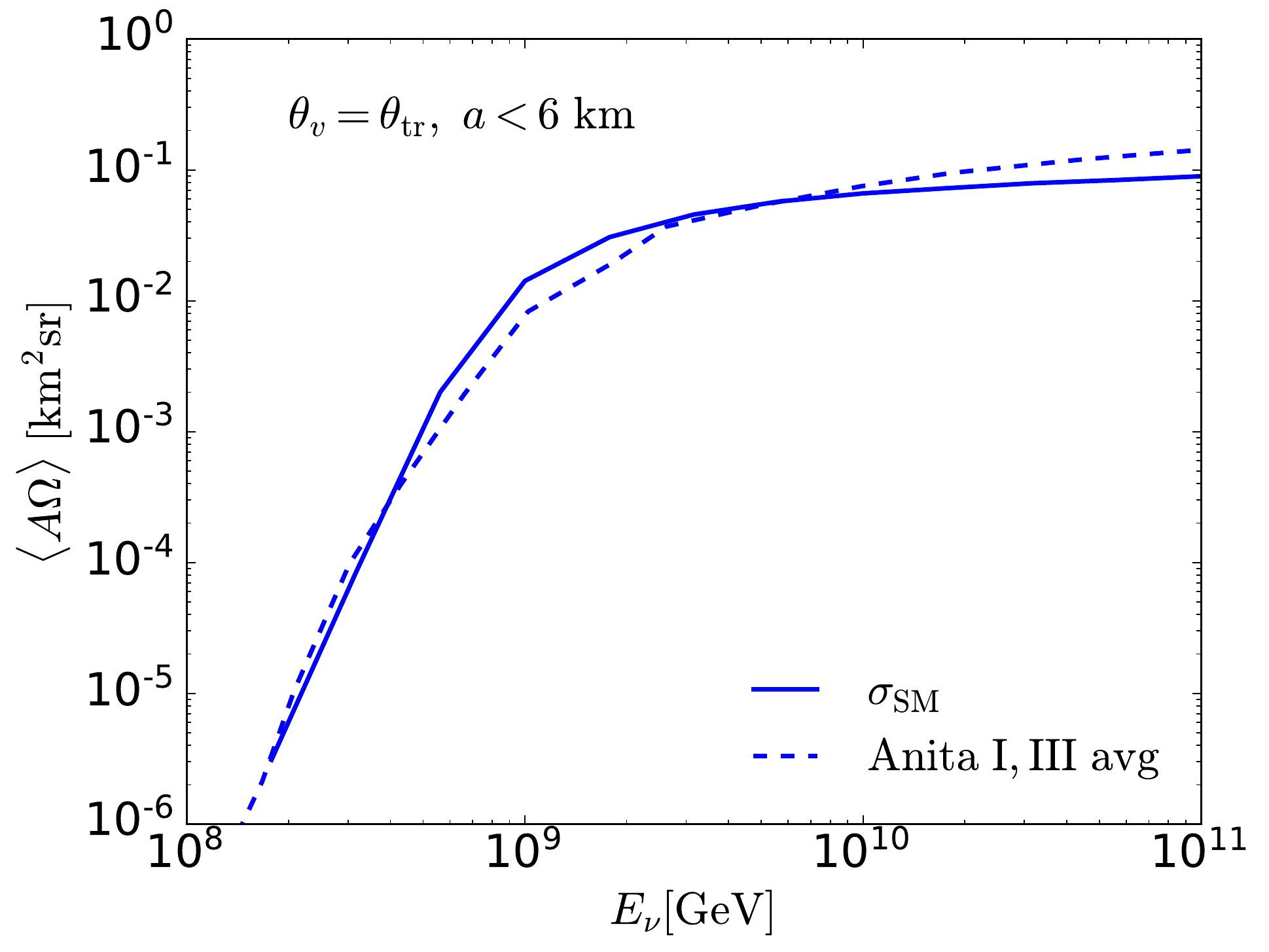}
\caption{{\it Upper}: The probability of a tau to exit for fixed $E_{\nu_\tau}=10^8$,
$10^9$, 10$^{10}$ and 10$^{11}$ GeV in the standard model, as a function of elevation angle $\beta_{\rm tr}$. The ALLM photonuclear energy loss is used.
{\it Lower}: The standard model effective aperture compared with the mean effective aperture of Anita I,III \cite{Romero-Wolf:2018zxt} for upward-going tau air showers from incident $\nu_\tau$'s.}
\label{fig:standard}
\end{figure}

The upper panel of Fig. \ref{fig:standard} shows the tau exit probabilities for fixed energies as functions of the elevation angle of the exiting tau, $\beta_{\rm tr}$. The resulting effective aperture comes from the exit probabilities and the associated cumulative distribution functions for the exiting tau energies, together with the decay and detection probabilities. A comparison of our calculated effective aperture (solid line) and the mean ANITA I,III effective aperture from Ref. \cite{Romero-Wolf:2018zxt} (dashed line) in the lower panel of Fig. \ref{fig:standard} shows that our simplified model of the ANITA detection probability is reasonable. 

Smaller cross sections, either from saturation effects for standard model neutrinos or for sterile neutrinos with a suppressed cross section, change the angular dependence of $p_{\rm exit}(E_\tau | E_{\nu_\tau}
\,\beta_{\rm tr})$ because neutrino attenuation is reduced. We assume that the differential
cross section, relatively normalized with a suppression factor of $\epsilon_\nu$, is the same as for the standard model evaluated with CT14 parton distribution functions. The charged current cross section with $\epsilon_\nu=0.1$ is shown in Fig. \ref{fig:sigma}. Smaller cross sections also result in fewer tau regeneration effects as the neutrino propagates through long chord lengths in the Earth. The first interaction occurs deeper along the neutrino trajectory.

\begin{figure}[ht]
\includegraphics[width=\columnwidth]{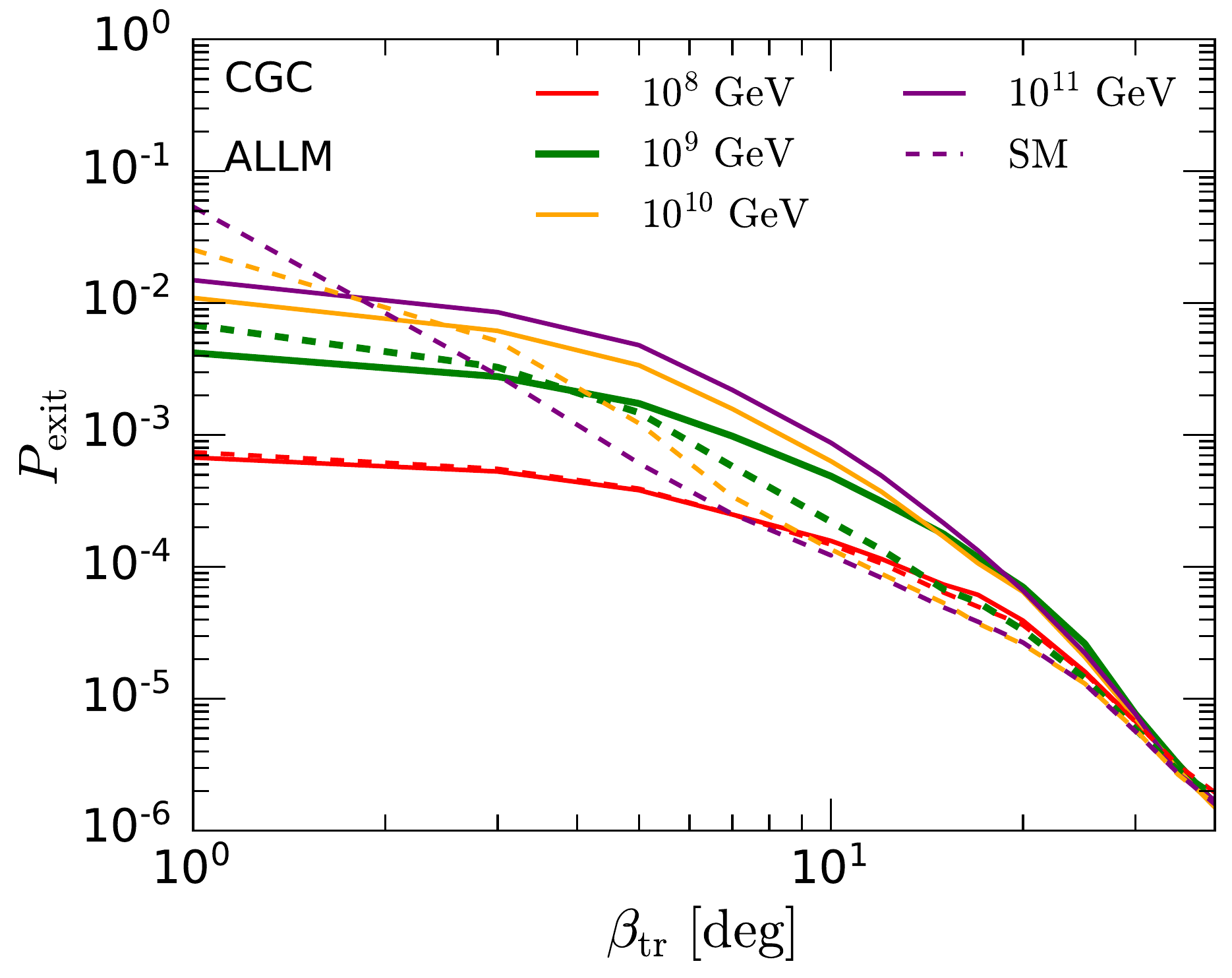}
\caption{The probability for a tau to exit for fixed $E_{\nu_\tau}=10^8$,
$10^9$, 10$^{10}$ and 10$^{11}$ GeV for the color glass condensate ultrahigh energy neutrino cross section extrapolation (solid lines) and for a parton distribution function evaluation of the cross section ($\sigma_{\rm SM}$, dashed lines), as in Fig. \ref{fig:standard}.}
\label{fig:prob-cgc}
\end{figure}

High energy extrapolations of the neutrino-nucleon cross section eventually face unitarity limits on the growth of the cross section. In the parton picture, the high
density of gluons at small parton momentum fraction $x$ is such that gluon recombination occurs, eventually saturating the cross section. One approach to handle
the saturation effects is the color glass condensate (CGC) 
formalism \cite{McLerran:1993ni,JalilianMarian:1996xn,McLerran:1998nk,Kovchegov:1996ty,Kovchegov:1997pc,Henley:2005ms}.
The high energy CGC extrapolation  of the neutrino cross section
is shown in Fig. \ref{fig:sigma} by the dashed red curve. This represents the strongest saturation effects
presented in Ref. \cite{Henley:2005ms}.
Fig. \ref{fig:prob-cgc} shows that the CGC extrapolation of the neutrino cross section
has some impact on the exit probability at large angles and at high energies. Overall, the exit probabilities still fall with increasing $\beta_{\rm tr}$ in the range of tens of degrees.

\begin{figure}[ht]
\includegraphics[width=\columnwidth]{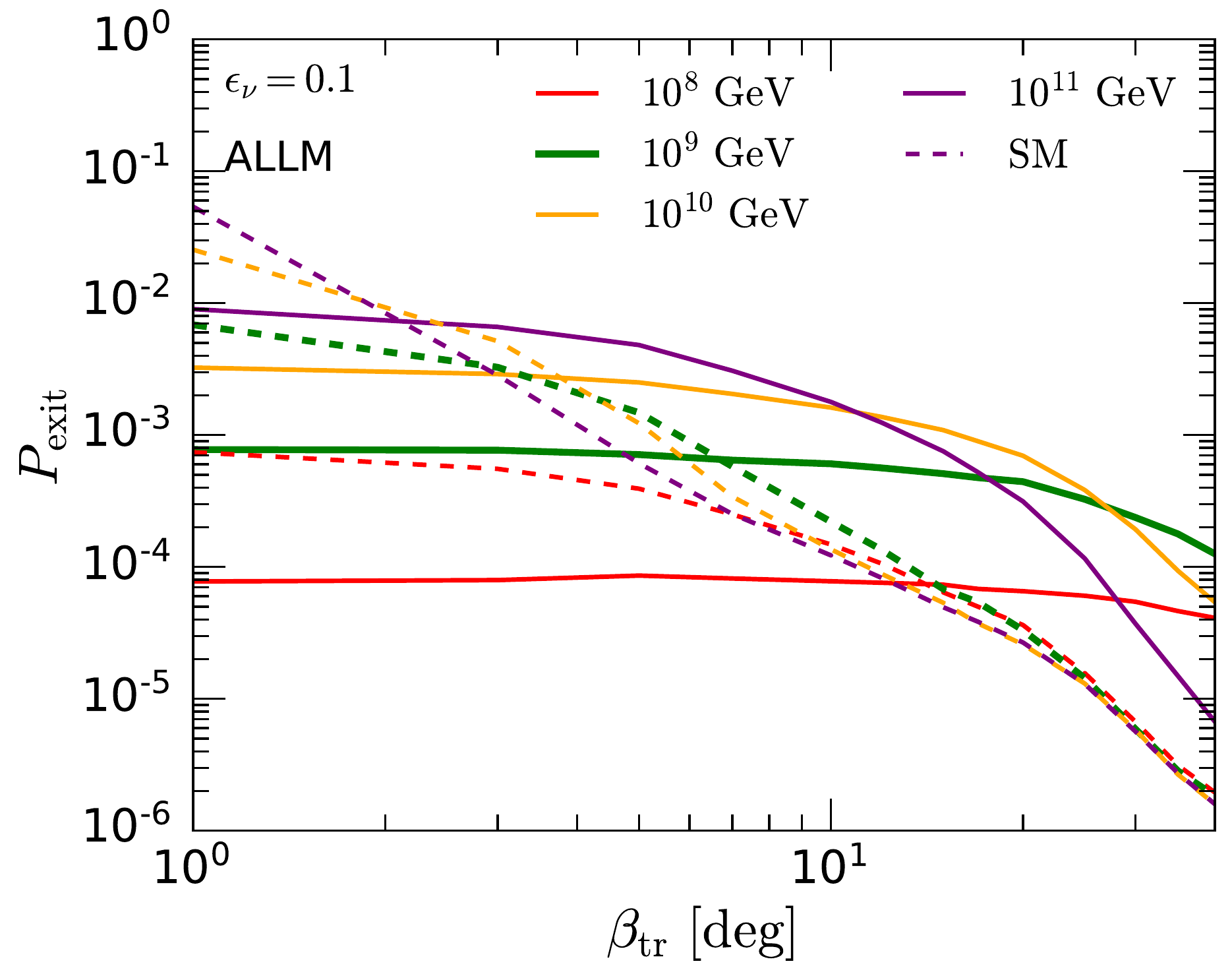}
\includegraphics[width=\columnwidth]{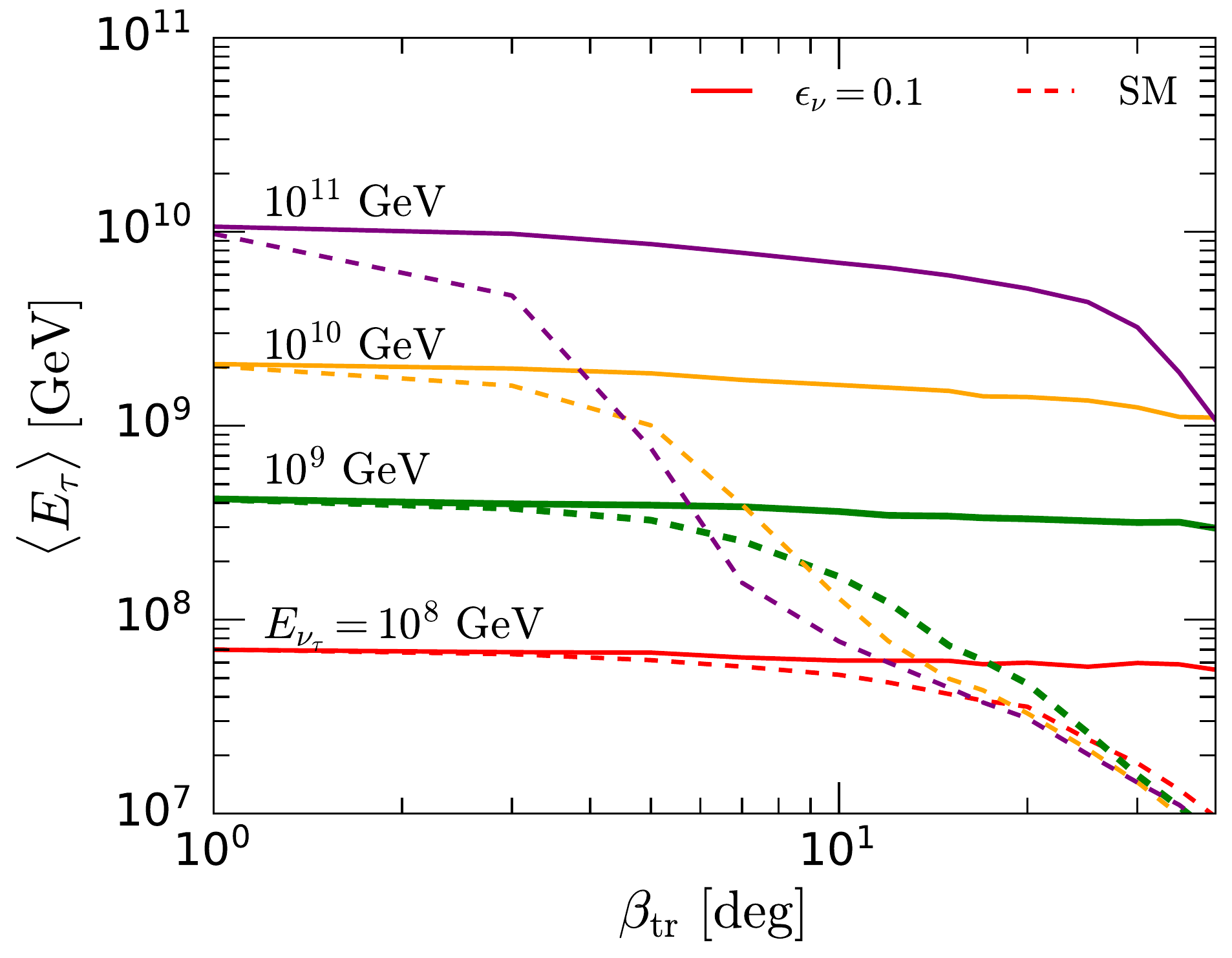}
        \caption{Upper: Probabilities for fixed sterile neutrinos energies as a function of elevation angle, for an $\epsilon_\nu=0.1$ sterile factor (solid lines) and for the standard model (dashed lines), using the ALLM model for photonuclear energy loss of the tau. Lower: The average energy of the emerging tau for sterile neutrinos with $\epsilon_\nu=0.1$ and the standard model.}
        \label{fig:prob-sterile}
\end{figure}

We now turn to the sterile neutrino cross section. 
Fig. \ref{fig:prob-sterile} shows the 
tau exit probabilities (upper) and average energy of the emerging taus (lower) for 
a sterile neutrino cross section $\sigma_{\nu_s N}=\epsilon_\nu\sigma_{\nu N}$,
with $\epsilon_\nu = 0.1$ (solid lines) and
for the standard model (dashed lines). We assume that the sterile neutrino interactions convert sterile neutrinos to tau neutrinos.

For small elevation angles (e.g., $\beta_{\rm tr}=1^\circ$), attenuation is not important. The smaller cross section for the sterile neutrino reduces the standard model tau exit probability by $\epsilon_\nu$. At larger angles, the exit probabilities for the sterile neutrino scenario do not fall as quickly as for the standard model because the sterile interaction length is longer. For $E_{\nu_s}=10^9$ GeV, the exit probability
for $\epsilon_\nu=0.1$
is more than an order of magnitude larger than for the standard model for $\beta_{\rm tr}=30^\circ$.

\begin{figure}
        \includegraphics[width=\columnwidth]{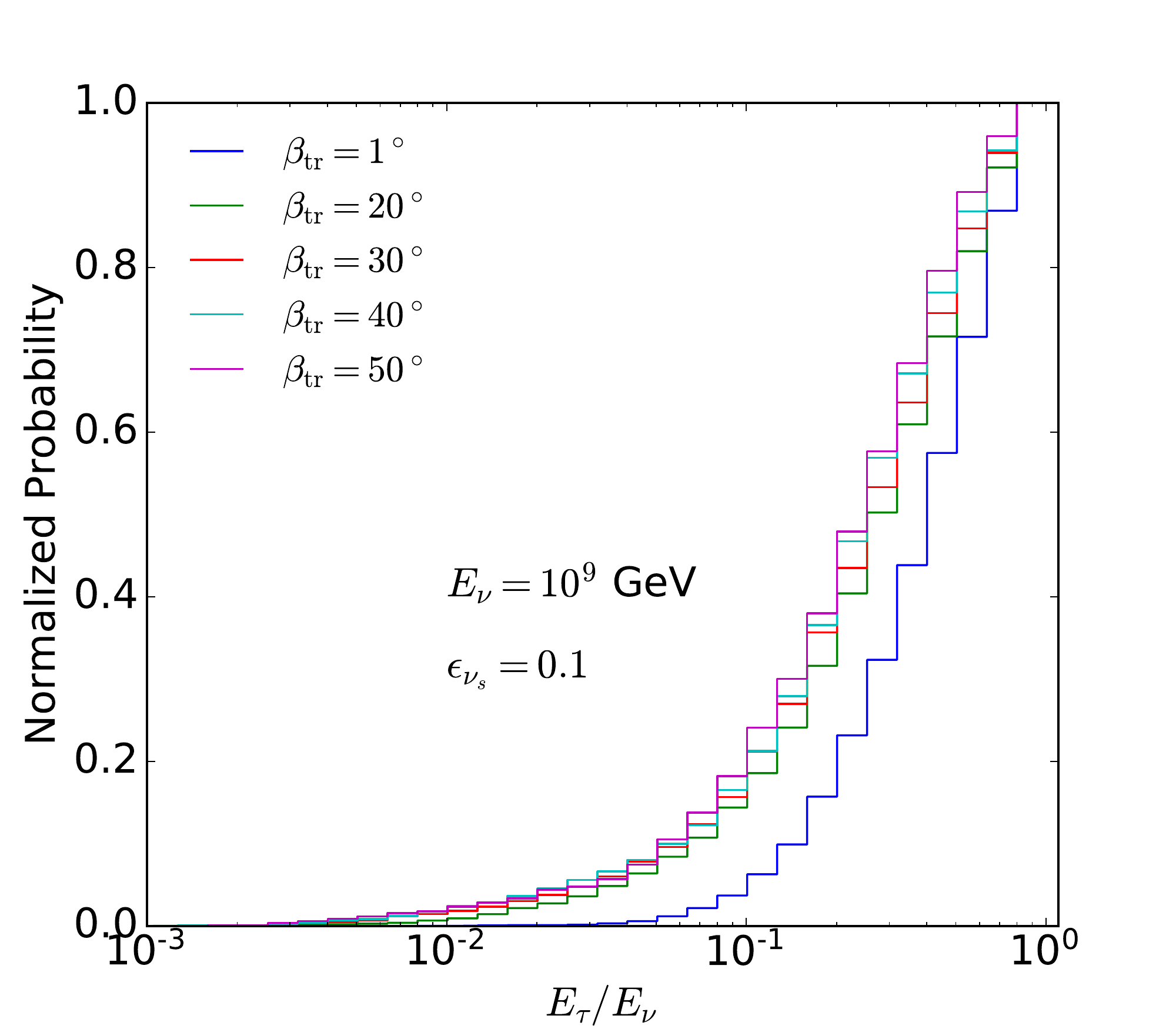}
        \caption{The cumulative distribution functions for several values of $\beta_{\rm tr}$ given $E_{\nu_s}=10^9$ GeV and 
        $\epsilon_\nu=0.1$.}
        \label{fig:cumulative-sterile}
\end{figure}

The lower panel in Fig. \ref{fig:prob-sterile} shows 
$\langle E_\tau\rangle$ as a function of $\beta_{\rm tr}$
for fixed $E_{\nu_\tau}$. The figure illustrates a second feature for $\epsilon_\nu=0.1$ that enhances tau shower detectability at large elevation angles. At $\beta_{\rm tr}=30 ^\circ$, for $E_{\nu_s}=10^9$ GeV and $\epsilon_\nu=0.1$, the average energy of the emerging tau is $\sim 3\times 10^8$ GeV, an energy more likely to be detected than the average energy of $\sim 2\times 10^7$
GeV of the standard model for the same  incident
neutrino energy and angle. The nearly constant $\langle E_\tau\rangle$ is evident from the cumulative distribution functions for exiting taus given a series of angles $\beta_{\rm tr}$ for a sterile neutrino energy $E_\nu=10^9$ GeV, shown in Fig. \ref{fig:cumulative-sterile}.

\begin{figure}
    \centering
    \includegraphics[width=\columnwidth]{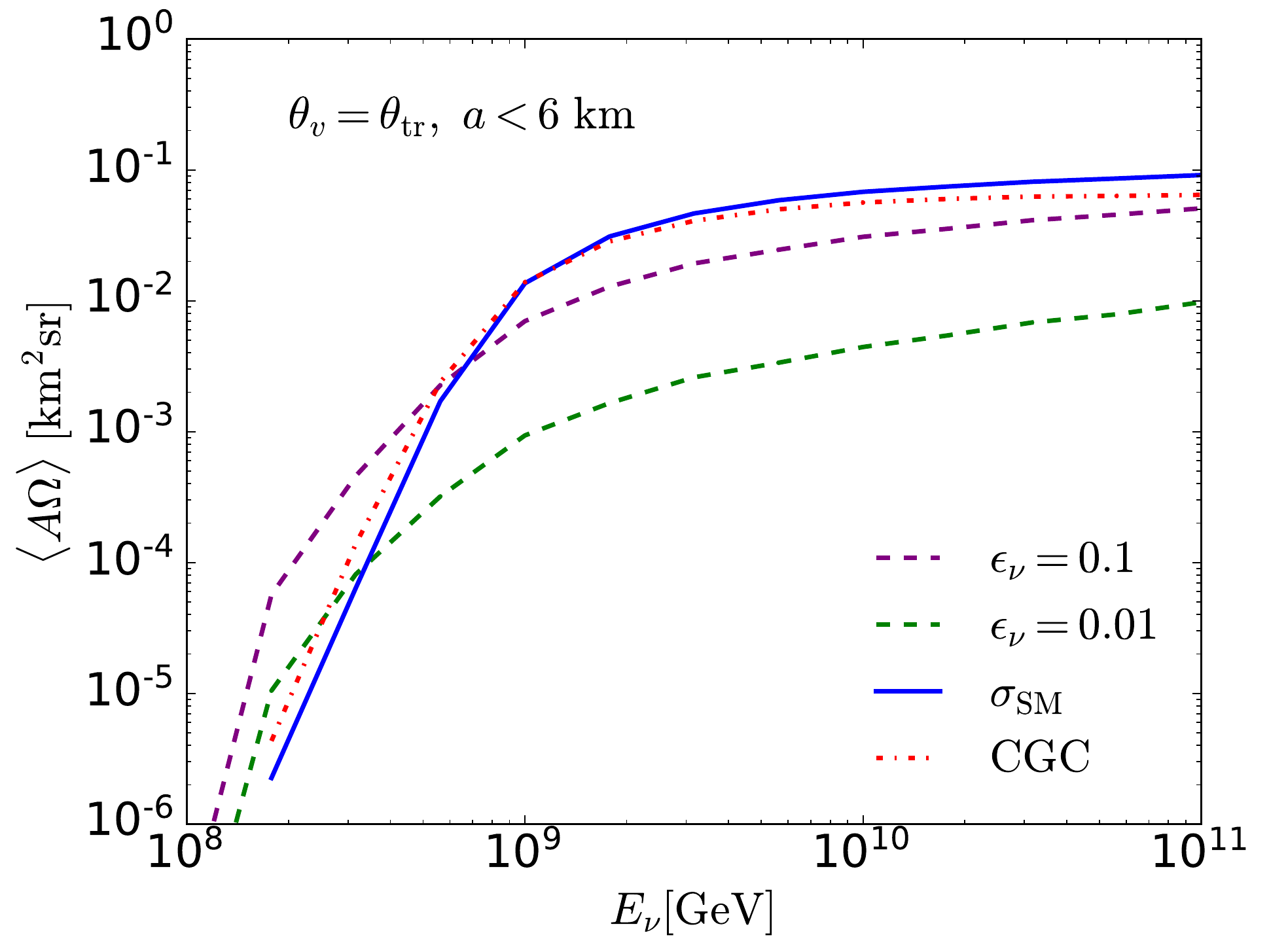}
    \caption{The effective aperture for standard model tau neutrinos and sterile neutrinos with $\sigma = \epsilon_\nu\sigma_{\rm SM}$, with 
    $\epsilon_\nu=0.1$ and the ALLM energy loss model. Also shown is the acceptance with a modified neutrino cross sections according to the color glass condensate model (CGC).}
    \label{fig:effective-aperture}
\end{figure}

Figure \ref{fig:effective-aperture} shows the effective aperture for standard model tau neutrinos with the CT14 cross section (solid line, labeled $\sigma_{\rm SM})$ and color glass condensate cross section (dot-dashed line, labeled CGC), and for sterile neutrinos with $\epsilon=0.1,\ 0.01$ (dashed lines). The CGC effective aperture is slightly larger than the standard model evaluation at low energies, and slightly lower than the standard model evaluation at high energies. 
The effective apertures for $\epsilon=0.1,\ 0.01$
are enhanced at low energies where the effective aperture increases with energy, but the maximum effective aperture is lower than for $\sigma_{\rm SM}$. 

\begin{figure}
    \centering
    \includegraphics[width=\columnwidth]{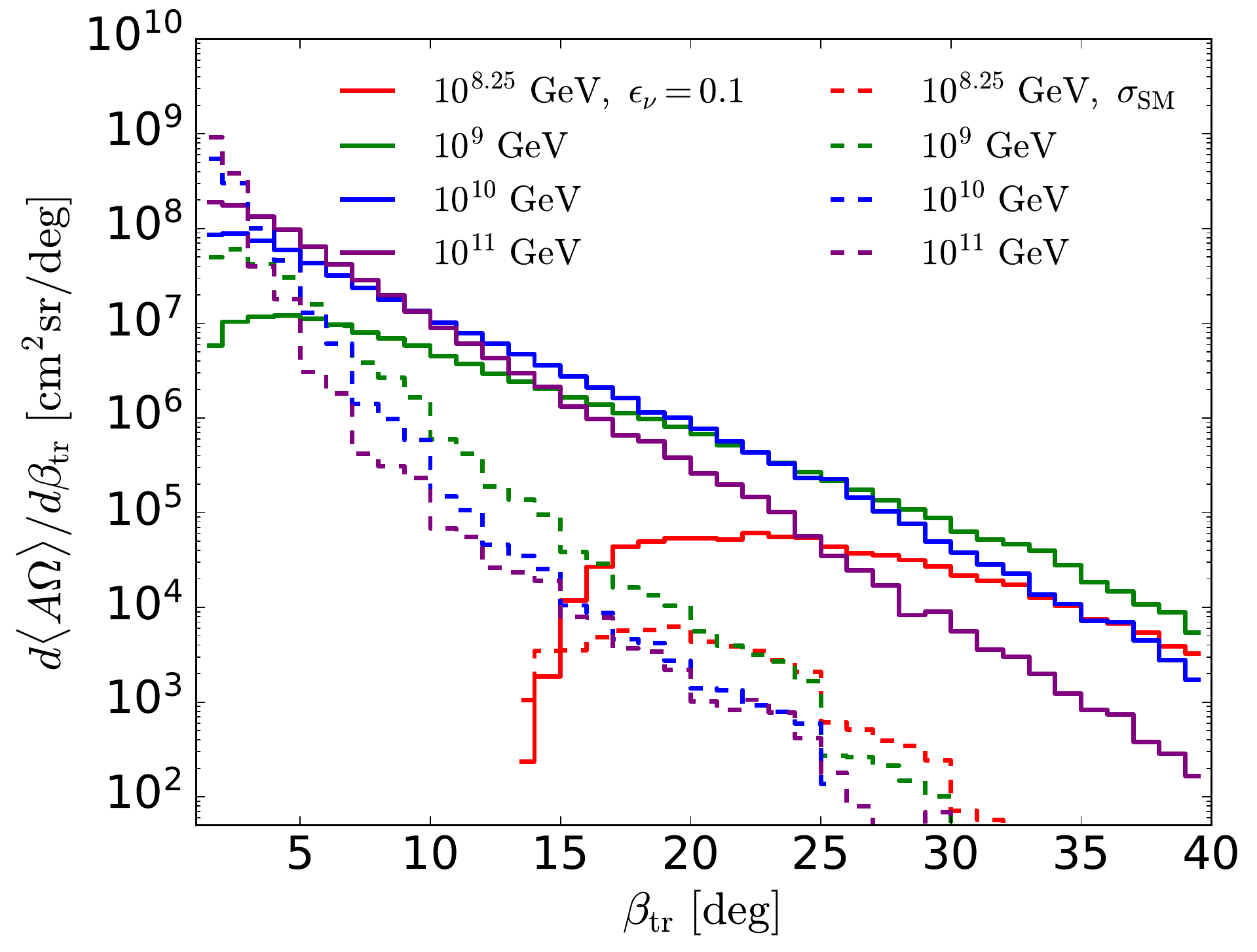}
    \caption{The differential effective aperture as a function of $\beta_{\rm tr}$ for standard model tau neutrinos (dashed) and sterile neutrinos (solid) with $\sigma = \epsilon_\nu\sigma_{\rm SM}$,  
    $\epsilon_\nu=0.1$ and the ALLM energy loss model.}
    \label{fig:dAOmega}
\end{figure}

\begin{figure}
    \centering
    \includegraphics[width=\columnwidth]{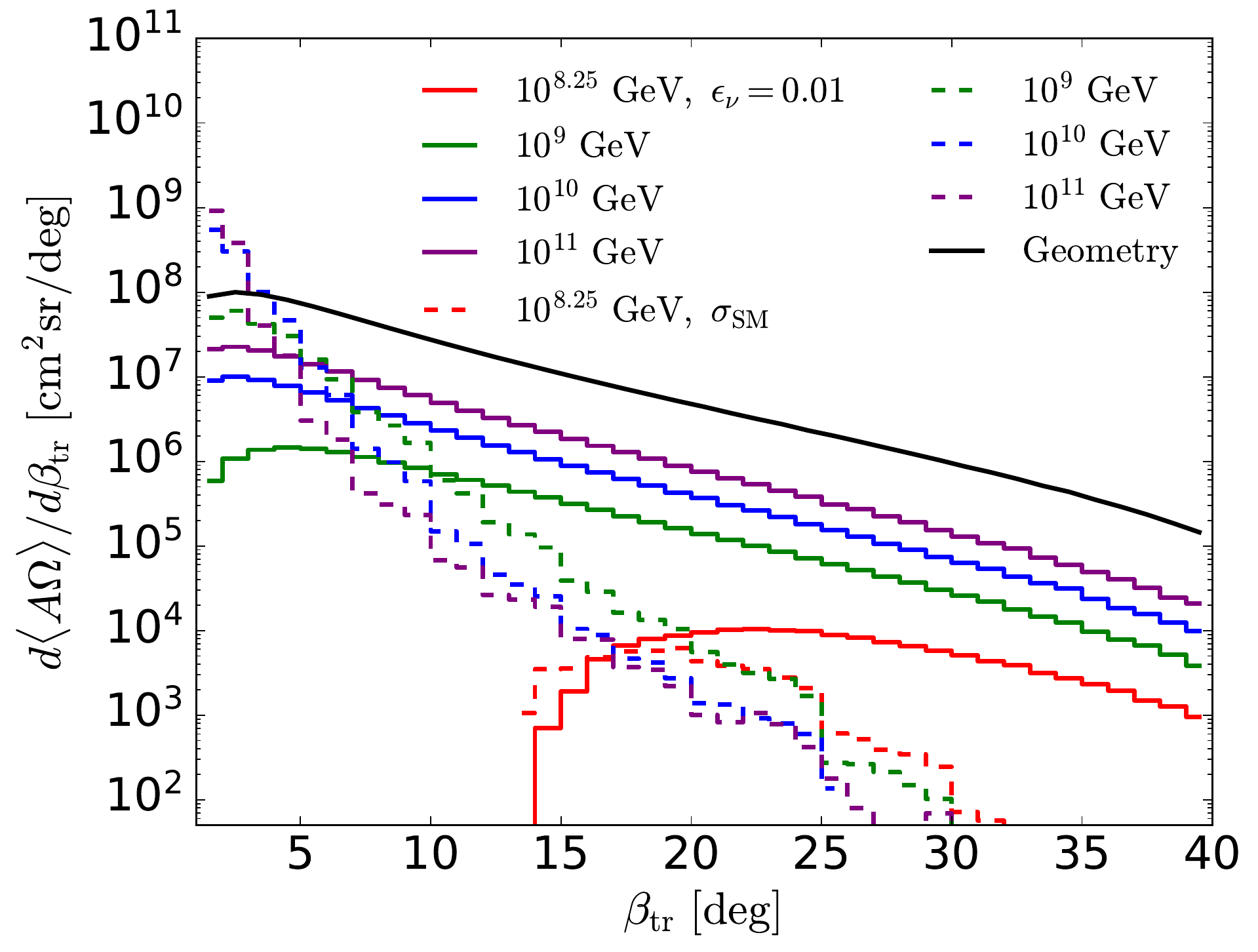}
    \caption{The differential effective aperture as a function of $\beta_{\rm tr}$ for standard model tau neutrinos (dashed) and sterile neutrinos (solid) with $\sigma = \epsilon_\nu\sigma_{\rm SM}$,
    $\epsilon_\nu=0.01$ and the ALLM energy loss model.
    The curve labeled ``Geometry" shows the rescaled differential aperture when $P_{\rm exit}=P_{\rm obs}=1$.}
    \label{fig:dAOmega-p01}
\end{figure}

The differential $\langle A\Omega\rangle$ as a function of $\beta_{\rm tr}$ is a useful diagnostic of the angular distribution of upward-going tau decay events \cite{Romero-Wolf:2018zxt}.
There is not a significant change to angular distributions of predicted events using the CGC extrapolation of the neutrino cross section compared to the standard evaluation, so we do not show it here.
For sterile neutrinos, the angular distribution changes, as shown in Figs. \ref{fig:dAOmega} and \ref{fig:dAOmega-p01} for $\epsilon_\nu=0.1$ and $\epsilon=0.01$, respectively.

The enhanced high $\beta_{\rm tr}$ distribution,
$d\langle A\Omega\rangle/d\beta_{\rm tr}$, for $\epsilon_\nu=0.1$ is shown with the solid lines in Fig. \ref{fig:dAOmega}. The standard model result is shown with the dashed lines.
For $\beta_{\rm tr}=30^\circ$, the differential effective aperture as a function of $\beta_{\rm tr}$ is 
$\sim 10^2-10^3$ times larger for sterile neutrinos with $\epsilon=0.1$ than for tau neutrinos.
Fig. \ref{fig:dAOmega} also shows that for $E_\nu=10^9$ GeV and $\beta_{\rm tr}=5^\circ$, the differential effective aperture is the same for standard model and sterile neutrinos with $\epsilon_\nu=0.1$, both larger by a factor of $\sim 100$ compared to the differential aperture for sterile neutrinos with $\epsilon_\nu=0.1$ at
$\beta_{\rm tr}=30^\circ$.

The larger differential aperture for small $\beta_{\rm tr}$ compared to $\beta_{\rm tr}\sim 30^\circ$ is qualitatively a consistent feature for all sterile neutrino cross sections, as we illustrate with $\epsilon=0.01$ in Fig. \ref{fig:dAOmega-p01}
with the solid histograms. For reference, we also show the standard model differential aperture, again with dashed histograms.

When $\epsilon_\nu=0.01$, except for $E_{\nu_s}\sim 10^{11}$ GeV, there is little angular dependence in $P_{\rm exit}$. For ANITA, with our model of the effective aperture, essentially all of the angular dependence is in the angle integrals over $\theta_E$ and in $\theta_{\rm Ch}^{\rm eff}$, once the shower threshold energy is reached. 

This effect can be understood by comparing the histograms in Fig. \ref{fig:dAOmega-p01} with the black line
labeled ``Geometry."
The solid black line comes from a rescaled geometric differential aperture, where $P_{\rm obs}=1$
and $P_{\rm exit}=1$ for all angles. For high sterile neutrino energies, $P_{\rm obs}\simeq 1$.
For low energies, at low angles, the
showers cannot be detected because of the long distance from tau exit point to ANITA. The distance from the exit point to ANITA for $\beta_{\rm tr}=1^\circ$ is $v=567$ km, while the decay length of the tau is $\gamma c\tau=5$ km for $E_\tau=10^8$ GeV. At high energies, the solid histograms in Fig. \ref{fig:dAOmega-p01}
increase with energy (for $\beta_{\rm tr}\gsim 5^\circ$) with a scaling that follows the energy dependence of the neutrino cross section, but the shape follows the geometric differential aperture.

 \begin{figure}
    \centering
    \includegraphics[width=\columnwidth]{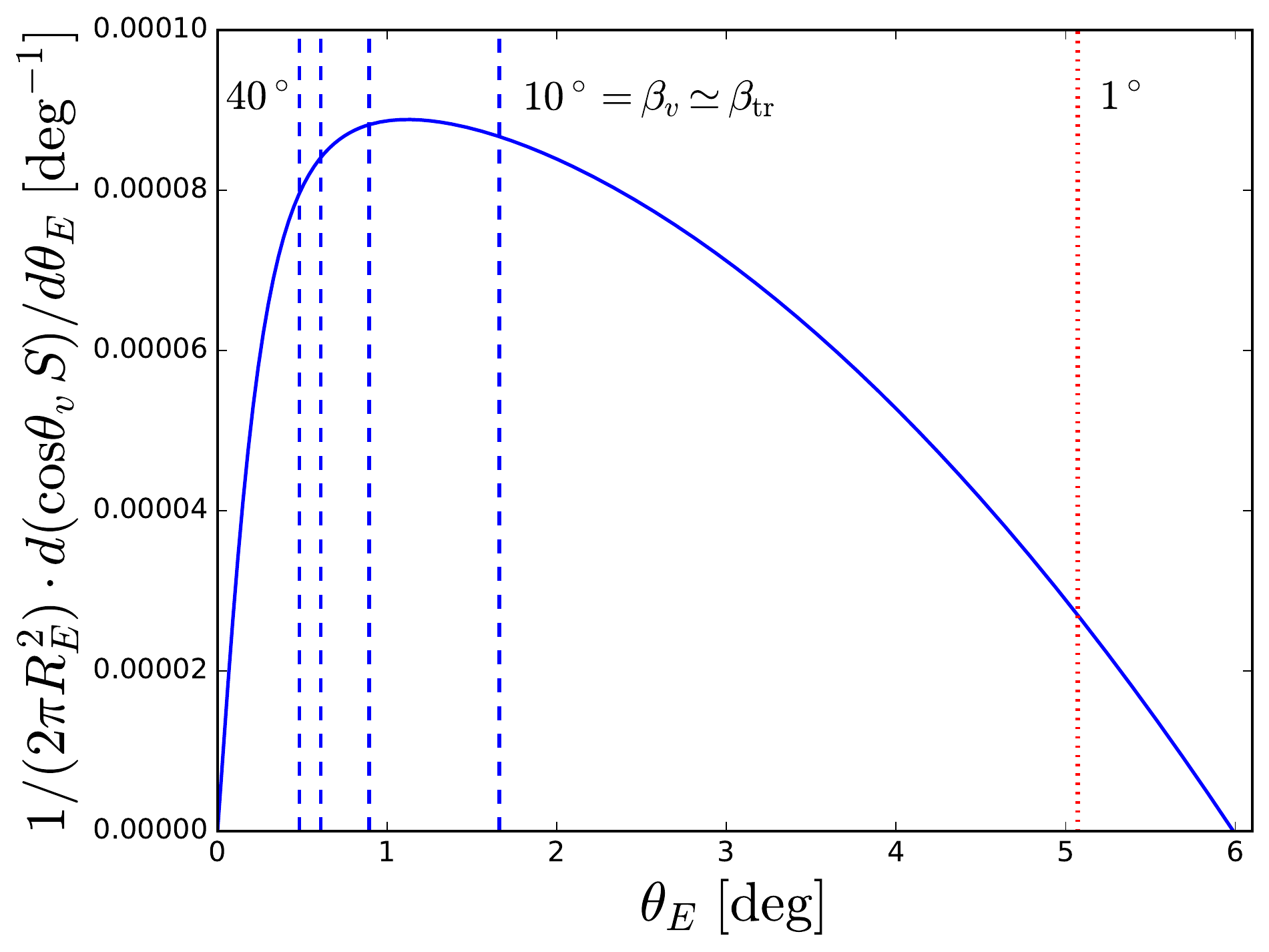}
    \caption{The differential effective area $S$ as a function of the co-latitude $\theta_E$ of a point on the surface in view, for $h=35$ km. The blue vertical dashed lines, from left to right, show the corresponding $\beta_{v}$
    for $\beta_v=40^\circ,\ 30^\circ,\ 20^\circ$ and $10^\circ$. The red dashed line corresponds to $\beta_v=1^\circ$.}
    \label{fig:dSdthetaE}
\end{figure}

To further illustrate the geometric effect,
Fig. \ref{fig:dSdthetaE} shows $({1/2\pi R_E^2})\, d(\cos\theta _v\, S)/d\theta_E$ where $dS$ is a patch of surface area in the viewing range of ANITA at co-latitude $\theta_E$,
as in eq. (\ref{eq:apereqn}).

 The blue curve starts at $\theta_E=0$, then increases as the annulus of area increases with $\theta_E$, then decreases as $\cos\theta_v\to 0$ as the angle relative to the local $\hat{n}$ goes to $90^\circ$. The vertical blue dashed lines mark where $\beta_{\rm tr}=40^\circ$, $30^\circ$, $20^\circ$ and $10^\circ$ are located in terms of $\theta_E$. The red dotted line shows $\beta_{\rm tr}=1^\circ$. The interval $\beta_v=30^\circ-40^\circ$ contributes about 3\% of the integral under the curve in Fig. \ref{fig:dSdthetaE}. The interval  $\beta_v=0^\circ-10^\circ$
 makes up $\sim 64\%$ of the integral.
 This geometric effect cannot be overcome by modifications of sterile neutrino cross section and/or a large isotropic sterile neutrino flux. 
 
 A flaring point source of neutrinos could be responsible for the ANITA events as noted in, e.g.
 Refs. \cite{Cherry:2018rxj,Cline:2019snp}. Other scenarios to produced
 anisotropies must overcome the geometric
 factor. 
For the angular range of the unusual events, given that 
$\beta_{\rm tr}\simeq \beta_v=25^\circ-35^\circ$ contributes $\sim 5\%$ to
the geometric surface area, an anisotropy must have more than a factor of 20 in the angular range of the ANITA
unusual events compared to skimming angles with $\beta_{\rm tr}\simeq \beta_v<10^\circ$.

\section{Discussion}

Are there circumstances where tau decays in the atmosphere can produce more upward air shower events at ANITA for $\beta_{\rm tr}\simeq 30^\circ$ than for $\beta_{\rm tr}\simeq 5^\circ$? Figures \ref{fig:dAOmega} and \ref{fig:dAOmega-p01} give a hint of the potential for relatively low energy $\sim 10^8$ GeV sterile neutrinos or other non-standard model neutral particles to produce large $\beta_{\rm tr}$ signals compared to small $\beta_{\rm tr}$. 

A key feature is that near the energy threshold of $\sim 10^8$ GeV for ANITA, large elevation angles are favored for detection. For diffuse neutrino fluxes that peak near ANITA's air shower threshold energy, the angular effect can be enhanced.

\begin{figure}[hbt]
    \centering
    \includegraphics[width=\columnwidth]{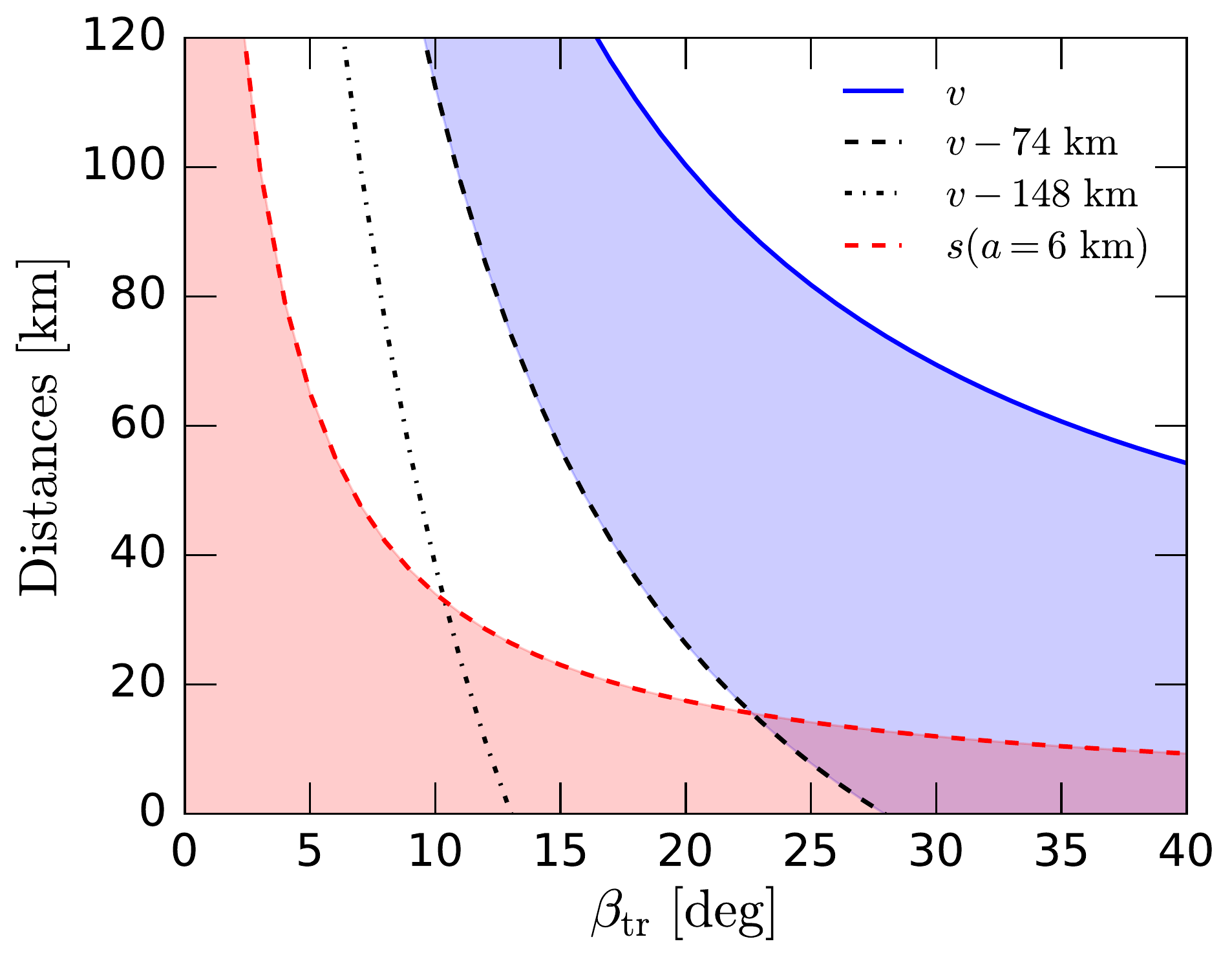}
    \caption{The pathlength $v$ of a trajectory with angle $\beta_{\rm tr}\simeq \beta_v$ with respect to the horizon, to an altitude of $a=35$ km.}
    \label{fig:pathlength-zoom}
\end{figure}

The step function for the detection probability $P_{\rm det}$ in eq. (\ref{eq:pdet}) requires
\begin{eqnarray}
\nonumber
    r_0 &=& v-s < 74\ {\rm km }\frac{E_{\rm shr}}{10^8\ {\rm GeV}}\\ v&-& 74 \ {\rm km }\frac{E_{\rm shr}}{10^8\ {\rm GeV}} < s\ .
\label{eq:r0}
\end{eqnarray}
The detection probability also requires that the altitude of the decay $a$ satisfy
\begin{equation}
    a<6\ {\rm km}\ .
    \label{eq:alt6}
\end{equation}
Figure \ref{fig:pathlength-zoom} shows the distances 
$v,\ s$ and $s(a=6\ {\rm km})$ in Eqs. (\ref{eq:r0}) and (\ref{eq:alt6}). The solid blue line shows the path length
$v$ between the tau exit point and ANITA
as a function of tau elevation angle at the exit point.
Since $E_{\rm shr}=0.98 E_\tau$ in the approximate aperture evaluation, we equate the shower energy and tau energy in the discussion here.
For $E_\tau = 10^8$ GeV, the difference between $v$ and the distance $s$ from the tau exit point to the decay must satisfy
$v-s<74$ km to be detectable. The blue dashed line in Fig. \ref{fig:pathlength-zoom} shows $s=v-74$ km as a function of $\beta_{\rm tr}$. For a shower energy of $10^8$ GeV to be detected, $s>v-74$ km.  The shaded blue region in the figure shows the allowed region for
$s$ given $E_\tau=10^8$ GeV.
For $\beta_{\rm tr}=1^\circ$, $v=567$ km, so a shower from a decay with $E_\tau=10^8$ GeV
($\gamma c\tau\simeq 5$ km) will be very rarely detected.
On the other hand, when $\beta_{\rm tr}=35^\circ$, $v=60.7$ km. All decay distances $s<v$ will satisfy the requirement in eq. (\ref{eq:r0}).

Our approximate effective aperture evaluation also requires the decay to occur below an altitude of $a=6$ km.
The path length $s$ at an altitude of 6 km as 
a function of $\beta_{\rm tr}$ is represented by the red dashed line in Fig. \ref{fig:pathlength-zoom}.
The shaded red region represents the allowed region of $s$ that satisfies the altitude requirement. With this model of the effective aperture, only values of $s$ in the overlapped red and blue shaded regions will be detected.
For $E_\tau \simeq E_{\rm shr}=10^8$ GeV,
Fig. \ref{fig:pathlength-zoom} shows that the taus can only be detected at angles $\beta_{\rm tr}\gsim 22^\circ$. This is the effect that is seen in the behavior of the differential aperture for $E_\nu=10^{8.25}$ GeV in Figs. \ref{fig:dAOmega} and
\ref{fig:dAOmega-p01}.

As the shower energy increases, the allowed region for $s$ also increases. The dot-dashed line in Fig. \ref{fig:pathlength-zoom}
shows the limit for $E_\tau=2\times 10^8$ GeV. The overlap of the region above the dot-dashed curve and below the red dashed curve is detectable. For this energy, $\beta_{\rm tr}\gsim 10^\circ$. Another factor of $2$ increase in energy moves the minimum $\beta_{\rm tr}$ to $\sim 5^\circ$.

The low energy $E_\tau \sim 10^8$ GeV air showers could, in principle, account for the large angle unusual ANITA events, but the effective aperture is small
for both standard model tau neutrinos and for sterile neutrinos. We illustrate the effect by evaluating ANITA's sensitivity to standard model tau neutrinos and sterile neutrinos. We use the effective aperture for $E_{\nu}=10^{8.25}\ {\rm GeV}=1.78\times 10^8$ GeV which shows an enhanced event rate for large elevation angles of the tau.  

We begin with standard model tau neutrinos. For this energy, $\langle
A\Omega\rangle\simeq 3.6\times 10^{-6}$ km$^2$sr for the
standard model.
ANITA's sensitivity to a tau neutrino energy squared scaled flux with standard model interactions, based on an exclusion at the 90\% unified confidence level in a decade of energy centered at $E_\nu= 10^{8.25}$ GeV for 115 days of ANITA I-IV flights, is 
\begin{eqnarray}
    \label{eq:sensitivity}
    {\rm Sensitivity}&=&\frac{2.44}{\ln(10)}
    \frac{1.78\times 10^8 \ {\rm GeV}}{\langle A
    \Omega\rangle \times 9.9\times 10^6\ {\rm s}}\\ 
    \nonumber
    &\simeq & 5.3\times 10^{-4}\ \frac{\rm GeV}{{\rm cm^2 s\, sr}}\ .
\end{eqnarray}
Standard model tau neutrino fluxes cannot be responsible for the ANITA unusual events as diffuse tau neutrino fluxes at this level
are already excluded by IceCube \cite{Aartsen:2018vtx} and Auger \cite{Zas:2017xdj}, as has already been emphasized recently by Romero-Wolf {\it et al.} in Ref. \cite{Romero-Wolf:2018zxt}. IceCube and Auger set upper bounds on the diffuse tau neutrino
differential flux (assuming equal fluxes neutrino flavors) in the range of  $E_\nu^2\Phi(E_\nu)\sim 10^{-8}-10^{-7}$ GeV/cm$^2$s\,sr for $E_{\nu}=10^8-10^{10}$ GeV.
 
The standard model tau neutrino effective aperture for ANITA rises quickly with energy, but as we have shown, this is accompanied by a larger
predicted number of events for small $\beta_{\rm tr}$ compared to the large elevation angles of the ANITA events. 
For $E_\nu=10^9$ GeV, $\langle A\Omega\rangle =1.4\times 10^{-2}$ km$^2$sr. Putting aside the question of angular dependence, ANITA's sensitivity to tau neutrinos is of order $\sim 8\times 10^{-7}$GeV/cm$^2$s\,sr for the decay of energy centered at $E_\nu=10^9$ GeV, still more than an order of magnitude higher than current limit from Auger of 
$\sim 2\times 10^{-8}$ GeV/cm$^2$s\,sr \cite{Zas:2017xdj}.

For sterile neutrinos, the larger effective apertures lead to better sensitivities for ANITA.  
With  $\epsilon_\nu=0.1\ (0.01)$, $\langle A\Omega(10^{8.25}\ {\rm GeV})\rangle\simeq 5.7\times 10^{-5}\ (1.1\times 10^{-5})$ km$^2$\,sr. For a sensitivity as defined in Eq. (\ref{eq:sensitivity}), ANITA's sensitivity to sterile neutrinos with $\epsilon_\nu=0.1\ (0.01)$ at $E_{\nu_s}=10^{8.25}$ GeV is 
$3.3\times 10^{-5}\ (1.7\times 10^{-4})$ GeV/cm$^2$s\,sr. For sterile neutrinos that oscillate with standard model neutrinos, the astrophysical tau flux is related to the sterile neutrino flux, so it is difficult to explain the unusual ANITA events with sterile neutrinos without over-predicting tau neutrino events in other detectors. In this paper, we are using the designation of sterile neutrino to denote a neutral particle with a cross section with nucleons to produce a tau that is smaller than the neutrino-nucleon cross section, so in principle, the flux of these particles does not have to be related to the diffuse cosmic neutrino flux.

One application of threshold energy enhancement of large angle events at ANITA is for monoenergetic sources. One example replaces the sterile neutrinos with $\chi$'s,
discussed in a 
recent paper by Hooper {\it et al.} \cite{Hooper:2019ytr}.
They propose that the unusual events at ANITA are Askaryan events from ultrahigh energy $\chi$ interactions, where supermassive dark matter $X_d\to \chi\chi$ decays in the galactic halo provide these mono-energetic, feebly interacting particles that have a cross section with nucleons that scales with the neutrino cross section: $\sigma_{\chi N}=\epsilon_\chi \sigma_{\nu N}$. Using a Navarro-Frenk-White density profile of dark matter and a local density normalization of $0.4$ GeV/cm$^3$, they find an integrated flux, averaged over 4$\pi$ steradians, of \cite{Hooper:2019ytr}:
\begin{equation}
    F_\chi\simeq \frac{52}{\rm km^2yr\,sr}
    \times \Biggl( \frac{2\times 10^{26}\ {\rm s}}{\tau_{X_d}} \Biggr)\times
    \Biggl( \frac{ 10^{11}\ {\rm GeV}}{m_{X_d}} \Biggr)\ ,
\end{equation}
in terms of the supermassive dark matter mass and lifetime. They constrain $m_{X_d}$ and $\tau_{X_d}$ based on an observing time of 115 days of flight of ANITA I-IV, assuming no unusual events are found with ANITA IV.
Hooper {\it et al.} find that the superheavy dark matter mass must be $m_{X_d}\gsim 1-2\times 10^{10}$ GeV for small $\epsilon_\chi$ \cite{Hooper:2019ytr} if the unusual events are Askaryan events. 
Our effective aperture can be carried over by substituting $E_\nu\to E_\chi$ and $\epsilon_\nu\to \epsilon_\chi$.
If we set $\epsilon_\chi=0.1 (0.01)$ and $E_\chi =10^{10}$ GeV,
two Askaryan events for ANITA in 115 days corresponds to $\sim 0.01 (0.02)$ shower events in the same time period.

If the two unusual ANITA events are not Askaryan events but instead from upward air showers from $\chi$ interactions with nucleons to produce $\tau$'s, ANITA is
sensitive to a different region of ($m_{X_d},\tau_{X_d}$
parameter space.
For $E_\chi=10^{8.25}$ GeV from the two body decay of $X_d$,  $m_{X_d}=3.56\times 10^8$.  For two events, the
integrated flux is determined to be $F_\chi \simeq 1.1\times 10^5/{\rm km^2\, yr\, sr}$
for $\epsilon_\chi=0.1$ and $F_\chi \simeq 5.8\times 10^5/({\rm km^2 yr\, sr}$ for
$\epsilon_\chi=0.01$. For smaller fractions $\epsilon_\chi$, attenuation in the Earth does not play a role for $\chi$ propagation, so for $E_\chi=10^{8.25}$ GeV,
\begin{equation}
F_\chi =   \frac{5.8\times 10^5}{\rm km^2 yr\, sr} \frac{0.01}{\epsilon_\chi} \ 
\end{equation}
to account for two ANITA events in 115 days. 

Events from $\chi$ induced showers in the ice with $E_\chi=10^{8.25}$ GeV are below ANITA's Askaryan energy threshold, so they would not be seen in Askaryan events.
However, IceCube should see these high energy events.
The number of downward IceCube events from $\chi$ interactions with $\epsilon_\chi=0.1$ is estimated to be 
\cite{Hooper:2019ytr}
\begin{equation}
    N\simeq 31/{\rm yr} \quad{\rm for}\ \epsilon_\chi=0.1\, ,
\end{equation}
for these input parameters, using $V\Delta\Omega = 1$ km$^3\times 2\pi$ sr. For smaller
$\epsilon_\chi$
\begin{equation}
    N\simeq 16/{\rm yr}\quad{\rm for}\  \epsilon_\chi<0.01\, .
\end{equation}
The number of events does not depend on $\epsilon _\chi$ for small values because $F_\chi$ scales as $\epsilon_\chi^{-1}$ and the $\chi N$ cross section scales with $\epsilon_\chi$. When attenuation in the Earth is negligible for $\chi$ transmission, the upward event rate should be independent of $\epsilon_\chi$.

The $\chi\to \tau$ events in IceCube 
would look like
$\nu_\tau$ production of $\tau$'s in the detector: there will be a hadronic shower with an associated tau.
For $E_\chi=10^{8.25}$ GeV, the average hadronic shower energy is $\sim 36$ PeV, and the average tau energy, $\sim 1.4\times 10^8$ GeV. Since $\gamma c\tau>5$ km for the tau at this energy, the tau will look like a muon with energy loss that is a factor of $\sim m_\mu/m_\tau \sim 0.05$ relative to a muon. Thus, the taus associated with the $\sim 36$ PeV showers would appear to be $\sim 8$ PeV ``muons." Muon-like tracks in the PeV energy range associated with cascades in the tens of PeV energy range, at a level of 16-31 events per year, are not observed, so this mass is excluded for $X_d$ particles. Larger masses don't favor large elevation angles, and smaller masses put $E_\chi$ below detection thresholds.

\section{Conclusions}

We have made an evaluation of the effective aperture for ANITA using a simplified model of the detection probability that is in reasonably good agreement with other results \cite{Romero-Wolf:2018zxt} for standard model tau neutrinos with  perturbative neutrino-nucleon cross sections. A modified neutrino cross section, with  high energy saturation effects modeled by color glass condensate suppression, does not have a large impact on the effective aperture. With a focus on the angular distribution of the effective sensitivity, we conclude that an isotropic flux of tau neutrinos cannot account for the large elevation angle ANITA event in the absence of skimming events. We concur with the authors of Ref. 
\cite{Romero-Wolf:2018zxt}.

Our quantitative evaluation is extended to particles with cross sections suppressed by a factor of $\epsilon_\nu$ relative to the standard model. We presented results
for $\epsilon_\nu=0.1$ and 0.01. The Monte Carlo simulation results 
for $\epsilon_\nu=0.01$ can be simply rescaled
by a factor of $\epsilon_\nu/0.01$ for $\epsilon_\nu<0.01$, since we have demonstrated that the main angular effect is geometric, not related to attenuation in the Earth or detection if the energy is above $\sim 10^9$ GeV and $\beta_{\rm tr}\gsim 5^\circ$.
Our results are more generally applicable to neutral particles incident on the Earth with feeble interactions that produce taus, as we showed with the supermassive dark matter model of Ref. \cite{Hooper:2019ytr}.

We showed that near threshold for ANITA, the geometric effects of the detection condition favor large elevation angles for the tau, but the effective aperture is small. At higher energies, the small angles are more important to the overall aperture. Our conclusion is that 
even with suppressed cross sections for sterile neutrinos and other feebly interacting particles, tau decay air showers cannot account for the ANITA events and be reconciled with IceCube and/or Auger limits.
Anisotropic sources that enhance event rates in the  $\beta_{\rm tr}\simeq 25^\circ-35^\circ$ degree range must account for a factor of $\sim 20$ compared to the $0^\circ-10^\circ$ degree range based on geometric effects alone.

\acknowledgements

Thanks to Luis A. Anchordoqui, Atri Bhattacharya, John F. Krizmanic, Angela V. Olinto, and Tonia M. Venters for discussions. This work was supported in part by US Department of Energy grants DE-SC-0010113 DE-FG02-13ER41976 and DE-SC-0009913 and NASA grant 80NSSC18K0246.



\end{document}